\documentclass[epj]{svjour}
\usepackage{graphicx}
\usepackage{amssymb}
\usepackage{lineno}
\usepackage{lineno}
\begin{document}
%
\titlerunning{Study of collective flows of protons...}
\title{Study of collective flows of protons and
$\pi^{-}$ -mesons  in p+C, Ta and He+Li, C collisions at momenta of
4.2, 4.5 and 10 AGeV/c }
\author{L. Chkhaidze\inst{1}
\thanks{\emph{corresponding author E-mail:} ichkhaidze@yahoo.com}
\and  G. Chlachidze\inst{2} \and
T. Djobava\inst{1} \and  A. Galoyan\inst{3} \and  L. Kharkhelauri\inst{1}
\and  R. Togoo\inst{4} \and  V. Uzhinsky\inst{5}
\thanks{\emph{corresponding author E-mail:} uzhinsky@jinr.ru}
}                     
%
%
\institute{High Energy Physics Institute of Tbilisi State
University, Georgia
 \and Fermi National Accelerator Laboratory, Batavia, Illinois 60510, USA
 \and Veksler and Baldin Laboratory of High Energy Physics, Joint
Institute for Nuclear Research, Dubna, Russia
\and Institute of Physics and Technology of the Mongolian Acad. Sci.,
Ulan Bator, Mongolia
\and Laboratory of Information Technologies, Joint Institute for Nuclear
Research, Dubna, Russia
}
\date{Received: date / Revised version: date}
%
\abstract{Collective flows of protons and $\pi^{-}$-mesons are studied at the
momenta of 4.2, 4.5 and 10 AGeV/c for p+C, Ta and He+Li, C interactions. The
data were obtained from the streamer chamber (SKM-200-GIBS) and from the Propane
Bubble Chamber (PBC-500) systems utilized at JINR. A method of Danielewicz
and Odyniec has been employed in determining a directed transverse flow of
particles. The values of the transverse flow parameter and the strength of
the anisotropic emission were defined for each interacting nuclear pair. It is
found that the directed flows of protons and pions decrease with increasing
 the energy and the mass numbers of colliding nucleus pairs. The
$\pi^{-}$-meson and  proton flows exhibit opposite directions in all studied
interactions, and the flows of protons are directed in the reaction plane.
The Ultra-relativistic Quantum Molecular Dynamical Model (UrQMD) coupled with 
the Statistical Multi-fragmentation Model (SMM), satisfactorily describes
 the obtained experimental results.
} 
\maketitle
\section{Introduction}

One of the central goals of the high-energy heavy-ion collision research is
a determination of the properties of nuclear matter at densities and temperatures
higher than that in the ground-state nuclei. Asymmetry of produced particle
emission relative to the reaction plane observed in nucleus-nucleus interactions
is explained by the collective flows of the particles. Theoretically, those flows
can be linked to the fundamental properties of nuclear matter and, in particular,
to the equation of state (EOS) \cite{R1}. Two types of asymmetries were identified.
The former is a directed flow \cite{R2} in the reaction plane, associated with the matter
"bouncing-off" within the hot participant region of overlap between colliding
nuclei. The latter is a squeeze-out \cite{R3} of the hot matter directed perpendicular
to the reaction plane within the participant region. When energy increases into
ultra-relativistic values the squeeze-out turns into an in-plane elliptic flow.

Many different methods have been proposed for studying the flow
effects in relativistic nuclear collisions, of which the most
commonly employed ones are the directed transverse momentum
analysis technique developed by Da\-nielewicz and Odyniec
\cite{R4} and the method of the Fou\-rier expansion for
azimuthal distributions proposed by Demoulins et al. and
Voloshin and Zhang \cite{R5}. Quantitatively, the anisotropic flow
is characterized by coefficients in the Fourier expansion of
the azimuthal dependence of the invariant yield of particles
relative to the reaction plane \cite{R6}. A first
coefficient, $v_{1}$, is usually called a directed flow
parameter, and a second coefficient, $v_{2}$, is called
an elliptic flow parameter. To improve anisotropic flow
measurements, advanced methods based on multi-particle
correlations (cumulants) have been developed to suppress
non-flow contribution, which are not related to the initial
geometry \cite{R7}. Most of the data below 4 AGeV in the
literature were, in fact, analyzed by the method of Danielewicz
and Odyniec \cite{R4}. The advantage of this method is that it
can be employed even at small statistics, which is typical for
film detectors. That is why we have chosen the method of
Danielewicz for the analysis of our data from film detectors in
order to investigate the directed flow of protons and
$\pi^{-}$-mesons in the colliding systems p+C, Ta and He+Li, C.

The collective flow of charged particles has been first
observed at the Bevalac by the Plastic Ball and Streamer
Chamber Collaborations. The flow continued to be explored at
Berkeley and at GSI, and further at AGS and at CERN/SPS
accelerators. By now, the collective flow effects were
investigated in nucleus-nucleus interactions over a wide range
of energies, from hundreds of MeV up to 5.02 TeV
\cite{U1}--\cite{R12}. Measurements of multi-particle azimuthal
correlations (cumulants) for charged particles in Pb+Pb
collisions at $\sqrt{s_{NN}}$ = 2.76 TeV and in p+Pb collisions
at $\sqrt{s_{NN}}$ = 5.02 TeV have been carried out at ALICE,
ATLAS and CMS LHC experiments. They help address the question
of whether there is an evidence for global, flow-like, azimuthal
correlations in these systems.

In order to study the properties of nucleus-nucleus
interactions, the collective flows of protons, pions and
$\Lambda$ hyperons were previously investigated
\cite{R13}--\cite{R16} by the authors of the present paper at
beam energies of 3.4 and 3.7 AGeV. According to the study, the
flow parameters of protons and negative pions increase at an
increase of the mass of projectile and target nuclei
($(A_{P}+A_{T})^{1/2}$) in $^2$H + C, He + C, C + C, C + Ne, $^2$H + Ta, He + Ta,
C + Cu, and C + Ta collisions. It was interesting to study
characteristics of the collective flow in nucleon-nucleus
interactions also.

It is worth to mention that the values of the elliptic flow
excitation function $v_{2}$ obtained by us for protons
correspond to quite an interesting energy region. According to the
investigations of Au+Au collisions at AGS \cite{R17}, an
evolution from a negative ($v_{2}<$ 0) to positive ($v_{2}>$ 0)
elliptic flow was observed at an energy interval of 2.0 $\leq$
Ebeam $\leq$  8.0 GeV/nucleon and an apparent transition energy
$E_{tr} \sim$ 4 GeV/nucleon was pointed. Therefore, our results
are also interesting for enrichment of the existing results in
the above mentioned energy region.

The collective flows are well established in collisions of heavy nuclei,
while the information is very limited for  interactions of light and medium
projectile nuclei with various target nuclei. The results obtained in
our paper will bring a new light on the nature of the flows.

In this paper, we present the collective flow analysis results of protons and $\pi^{-}$
-mesons for different projectile-target combinations at the laboratory momenta
of 4.2, 4.5 and 10 AGeV/c. The data were obtained from the experimental setups of JINR,
Dubna. Moreover, the characteristics of protons and $\pi^{-}$ -mesons, produced
in the collisions, were determined and provided for comparison at different
energies. The experimental results are compared with the predictions
of the Ultra-relativistic Quantum Molecular Dynamics Model (UrQMD)
\cite{R18} in combination with the Statistical Multi-fragmenta\-tion Model (SMM) \cite{R19}. 

\section{ Experimental data}
The data were obtained from the SKM-200-GIBS streamer chamber  

The SKM-200-GIBS experiment is based on a 2 m streamer chamber
placed in the magnetic field of 0.8 T and on a triggering
system. An inelastic trigger was used to select the events. The
streamer chamber \cite{R20} was exposed by a beam of He nuclei
accelerated up to 4.5 AGeV/c in the JINR synchrophasotron. The
thickness of Li and C solid targets (in the form of a disc)
were 1.5 and 0.2 g/cm$^{2}$, correspondingly.  The analysis
produced 4020 events of He+Li and 2127 events of He+C collisions.         

The 2-meter long Propane Bubble Chamber (PBC-500) was placed in the magnetic field
of 1.5 T. The procedures for separating out the p+C collisions in propane
(C$_{3}$H$_{8}$) and the data processing details, including particle
identification and corrections, were described in \cite{R21}. The analysis
produced 5882 (10775 events in C$_{3}$H$_{8}$) and 16509 (28703 events
in C$_{3}$H$_{8}$) events of p+C interactions at the momentum of 4.2
and 10 GeV/c, correspondingly, and 2342 events of p+Ta  (at 10 GeV/c)
collisions. The protons with momentum p$<$150 MeV/c were not detected
150 200 MeV/c
within the PBC-500 as far as their track lengths are less than
2 mm (p+C interactions), and protons with p$<$200 MeV/c were
absorbed in Ta target plate (the detector biases). Thus,
the protons with momentum larger than 150 MeV/c were registered in
p+C interactions, and the protons with p$\geq$250 MeV/c in p+Ta
collisions.

The following factors were considered at estimations of systematic errors for the both set-ups:
the contamination of non-beam particles (contamination due to interactions of other projectile
nuclei with a charge less than required, Z$<$ Z$_{proj}$), the choice of the effective area for
registration of collisions, secondary interactions in the target, absorption of slow $\pi^{-}$
mesons in the target, contamination of $K^{\pm}$ and $\Sigma^{\pm}$ particles, visual
identification  of $\pi^{+}$ mesons, scanning losses (including tracks with a large angles at the plane of photography).
For SKM-200-GIBS experiment the corrections  due to the trigger, selecting
inelastic and central collisions, were added, while for PBC-500 the corrections for the selection of interactions with carbon nuclei from the collisions with  propane nuclei were taken into account.
The most important corrections were connected with the scanning losses which mainly reduce
the number of the observed secondary particles.
All the factors led to a systematic uncertainty of $\pi^-$ meson average multiplicity
on the level 2$\div$4.5 \%. For the analysis presented in our paper, we study kinematic properties of $\pi^-$ mesons and protons in narrow intervals of rapidity. In each rapidity bin the statistical errors were in 3$\div$4 times larger than the pointed systematic uncertainty.
Thus, we will not consider the systematic errors further.

A study of the collective flow phenomenon needs an "event-by-event" analysis,
which requires the exclusive analysis of each individual collision.
In this connection, there has been a necessity to perform identification
of $\pi^{+}$ mesons in order to separate $\pi^{+}$ mesons
from the positive charged particles. The procedure has been done as
described below.

In what concerning He+Li, C collisions, identification of positive
charged particles never has been done yet, and we have carried out our
effectively $\pi^{+}$ mesons identification for the first time.
It was assumed, that $\pi^{+}$ and $\pi^{-}$ mesons hit a given cell
of the plane (P$_L$, P$_T$) with equal probability in collisions of
isospin-symmetrical nuclei. Thus [13], it was assumed that a proton
yield in a given cell ($ij$) of the P$_L$--P$_T$ plane is presented as:
$$N^{prot}_{ij}=N^{pos}_{ij}-N^{neg}_{ij},$$
where $N^{pos}_{ij}$ is a number of positively charged tracks in the
cell, and $N^{neg}_{ij}$ is a number of negative charged tracks in
the cell. As known [13], the procedure works quite well for inclusive
distributions. In our case, we randomly choose how to treat a given
positive charged track. We treated the track as a proton one with a
probability $W_p=N^{prot}_{ij}/N^{pos}_{ij}$, and as a $\pi^+$ meson one
with the probability $W_{\pi^+}=N^{neg}_{ij}/N^{pos}_{ij}$.

Since $\pi^{+}$ mesons and protons in the Propane Bubble Chamber
(p+C, Ta collisions) have been  identified in the narrow interval of
momenta (up to 0.7 GeV/c), it was necessary the additional "identification"
of particles with higher momenta. It was done as follow:
The momentum distributions (spectra) of positive and negative particles was
divided into 10 intervals for p$>$0.750 GeV/c. The distributions were created
for isospin-symmetrical collisions (${\rm ^2H,\ ^4He,\ C+C}$). Using them,
the probabilities $W_p$ and $W_{\pi^+}$ were determined. According to the
probabilities, the positive charged tracks with p$>$0.750 GeV/c were
subdivided into "protons" and "$\pi^+$ mesons".
The results of the "identification" for He+Li, C and p+Ta collisions are
presented in Figs 1, 2 for inclusive distributions in a comparison with
the UrQMD model predictions. One can see that the spectrum of $\pi^{+}$
mesons well agrees with the experimental spectrum (distribution)
of $\pi^{-}$ mesons and with the simulated spectrum of $\pi^{+}$ mesons.

For the flow analysis, minimum four participant protons, N$_{part} \geq$ 4,
(the remaining protons after the "identification") were required in event.
In the experiment, the positively charged projectile
fragmentation products were identified as those characterized by the
momentum p$>$3.5 GeV/c (4.2 AGeV/c, 4.5 AGeV/c)  or p$>$7 GeV/c
(10 AGeV/c) and emission angle $\theta$ $<$ 3.5$^{\circ}$ and the target
fragmentation products -- by the momentum p $<$ 0.25 GeV/c in the target
rest frame.

\section{The directed flows of protons and $\pi^{-}$-mesons}
It has been investigated that the directed flow of protons and $\pi^{-}$ mesons
in the colliding systems p+C, Ta and He+Li, C employed a method of
Danielewicz and Odyniec \cite{R4}. The method relies on summation over
transverse momenta of selected particles in the events. In an experiment,
no determination of the impact parameter $b$ is possible. Therefore,
instead of $b$, a vector sum of transverse momenta of projectile and target
nuclear fragments or participant protons is used. The fragmentation regions
of projectile and target nuclei were outside the acceptance regions of
experimental setups in some experiments and, therefore, the reaction plane was defined by the second approach using the participant protons. The second approach is preferable also for the light nuclear systems, because the multiplicity of the participant
protons is larger than the number of detected nuclear fragments. Therefore, the participant protons were used in our study for the reaction plane determination.

The analysis was carried out in the laboratory system. To eliminate the
correlation of a particle with itself (autocorrelations) at a study of
proton flows, we estimated the reaction plane for each proton in the event
with a contribution of that proton removed from the definition of the
reaction plane. The reaction plane is spanned by the impact parameter vector
$\overrightarrow{b}$ and the beam axis. Within the transverse momentum method,
the direction of $\overrightarrow{b}$ is estimated event-by-event in terms of the vector constructed
from the proton transverse momenta:
\begin{equation}
\overrightarrow{Q_{j}}=\sum\limits_{i=1,i\not=j}\limits^{n}\omega_{i}
\overrightarrow{p_{i}\hspace{0.01cm}^{\perp}},
\end{equation}
where $\omega_i$ is a weight factor of i-th particle, $\omega_{i}$
= y$_{i}$ - y$_{c}$, y$_{i}$ is a rapidity of i-th participating proton,
and y$_{c}$ is an average rapidity of the participant protons
in an event \cite{R22}. The values of $y_c$ averaged over all events are
shown in Figs.~3, 4 by arrows.

The projection of a transverse momentum of a particle onto the estimated reaction plane is:
\begin{equation}
p_{xj}\hspace{0.01cm}^{\prime} = \{ \overrightarrow{{Q_{j}}}
\cdot
\overrightarrow{p_{j}\hspace{0.01cm}^{\perp}} /
\vert\overrightarrow{{{Q_{j}}}}\vert\}.
\end{equation}
The dependence of the projection on the rapidity $y$ was constructed for each
collision systems. 
For further analysis, the average transverse momentum
in the reaction plane, $<p_{xj}\hspace{0.01cm}^{\prime}(y)>$, is obtained
by averaging over all events in the corresponding intervals of rapidity.
Due to finite number of particles used in constructing of the vector
$\overrightarrow{Q}$ in (1), the estimated reaction plane fluctuates in
the azimuth around the direction of the true reaction plane. Because of
those fluctuations, the component $p_{x}$ of a particle momentum in
the true reaction plane is systematically larger than the component
$p_{x}\hspace{0.01cm}^{\prime}$ in the estimated plane:
\begin{equation}
\langle p_{x} \rangle   =
\langle p_{x}\hspace{0.01cm}^\prime \rangle / \langle
\rm cos(\Phi) \rangle ,
\end{equation}
where $\Phi $ is the angle between the true and estimated reaction planes.
The overall correction factor $k$ = 1 $/$ $\langle$\rm cos$(\Phi) \rangle$ is
a subject of a large uncertainty \cite{R4,R14}, especially for low multiplicity events.
In \cite{R4} the method for the definition of the correction factor has been proposed.
Each event is randomly divided into two almost equal sub-events and the vectors
$\overrightarrow{Q_{1}}$ and $\overrightarrow{Q_{2}}$ are constructed.
\par
Then the distribution of the azimuthal angle $\Phi$ between these two
vectors was obtained,
and the average $<$cos$\Phi>$ was determined.
$<$cos$\Phi>$ for all considered interactions are given in
Table 1. Figs. 3, 4 show dependencies of the average in-plane
transverse-momentum components on the rapidity for protons and
$\pi^{-}$ -mesons in p+C, Ta and He+Li, C collisions. The
values of the flow parameter $F$, which are the slopes of
$<p_{x}(y)>$ at its mid-rapidity cross-over, are given in Table
1. All presented error bars are only statistical ones.

The flow analysis was done also for the changed requirement
for the determination of the reaction plane, namely for N$_{part} \geq$ 3.
In this case the absolute values of the directed flow parameter $F$
for protons have been obtained 82.7 $\pm$ 4.3 MeV/c (N$_{event}$=1720 after cut)
for He+Li, C and 76.8 $\pm$ 4.9 MeV/c (N$_{event}$=1541 after cut)
for p+Ta collisions. One can see from Table 1, that the change
of the selection criterion of the events for the flow analysis
does not affect significantly the values of the flow parameters, while
the errors of F increases for the criterion N$_{part} \geq$ 4 due
to the decrease of statistics: for He+Li, C collisions 80.9 $\pm$ 16.2 MeV/c (N$_{event}$=786 after cut)
and for p+Ta collisions 76.1 $\pm$ 5.3 MeV/c (N$_{event}$=1141 after cut).

In view of the strong coupling between the nucleon and pion,
it is interesting to know, if pions also have a collective flow
behaviour in that presented nuclear systems and how is the pion
flow related to the nucleon one. An answer to this question is
given in Table 1. The pion's flow has been observed with
respect to the reaction plane determined by participant
protons.

One can see from the Table 1 and Figures 3, 4 that the
$\pi^{-}$ mesons flows are smaller than the proton ones. It is
apparent that the $\pi^{-}$ mesons and protons flow are in the
opposite in-plane directions, in the forward and backward
hemispheres, in all nuclear systems. The flow parameter
absolute values for $\pi^{-}$ mesons and protons decrease with
increasing of projectile momentum and of mass numbers $A_{T}$
of target nuclei, what is opposite the results in
nucleus-nucleus collisions ($^2$H + C, He + C, C + C,$^2$H +
Ta, He + Ta, and C + Ta collisions) obtained at the same energy
and on the same setup \cite{R14}, \cite{R16}. Interestingly,
the proton and $\pi^{-}$ mesons collective flow parameters in He+C
collisions of SKM-200-GIBS with the pure, solid thin targets
are comparable with those from collisions with C$_{3}$H$_{8}$
(PBC-500, our earlier results) \cite{R16}. Both results are in
agreement within errors.

The obtained experimental results from p+C, Ta and He+Li, C collisions are
compared with predictions of the UrQMD model coupled with the SMM model.
Detailed descriptions of the UrQMD and SMM  models can be found in \cite{R18} and \cite{R19},
correspondingly. The UrQMD model is a microscopic transport model based on the
covariant propagation of all had\-rons on classical trajectories in combination
with stochastic binary scatterings, colour string formation and resonance decay.
We added the SMM model to the UrQMD model to improve simulations of a baryon
production in target and projectile fragmentation regions. There is no a strong
subdivision between the participating protons and protons created at the de-excitation
of nuclear residuals. The SMM allows a simulation of the evaporated proton production,
and UrQMD generates the participating protons. Thus, we have in the model events
all processes presented at the experiment.

The UrQMD model is designed as a multi-purpose tool for studying a wide variety
of heavy ion related effects ranging from multi-fragmenta\-tion and collective
flow to particle productions and correlations in the energy range from SIS to RHIC.
In the used version of the UrQMD model (1.3)
with addition of the SMM, we consider  potential interactions between nucleons
and excitations of the residual nuclei. These allowed one to determine the
reaction plane by the participant protons.

The incorporation of UrQMD and SMM was done as it was proposed
in Ref.~\cite{Kh1,Kh2}. The UrQMD calculation was carried out
up to the transition time $t_{tr}=100$ fm/c. After that the
minimum spanning tree method \cite{Kh3} was employed for
determination of prefragments. It was assumed that two nucleons
form a cluster, if a distance between their centers is lower
than $R_{clus}=3$ fm. The choice of the parameters ($t_{tr}$
and $R_{clus}$) was considered in Ref.~\cite{Kh1,Kh2}. The
procedure allowed to determine mass numbers, charges, energies
and momenta of pre-fragments. The energy and momentum of a
prefragment were transformed to the prefragment rest frame
using the Lorentz transformation. The excitation energy of the
hot prefragment was calculated as a difference between the
binding energy of the pre-fragment and the binding  energies of
this pre-fragment in its ground state. Skyrme, Yukawa and
Coulomb potentials were used at a calculation of the binding
energy. The Pauli potential was not accounted.  A "hard"
Skyrme-type EoS with the nuclear incompressibility $K=300 MeV$
was used. Decay of the prefragments was simulated by SMM.

We have checked out that the UrQMD model without the potential interactions
(cascade mode of UrQMD) does not predict any anisotropic flows for
proton interactions with light nuclei. An inclusion of the potential interactions
leads to the flows, and allows one to estimate the excitation energy of
nuclear remnant. Evaporated protons, neutrons and nuclear fragments are
produced at a de-excitation of the remnants. The average kinetic energy
of the evaporated protons is below 30 MeV. An accounting of the protons
does not have an essential influence on our results.

20000 (4.2 GeV/c) and 60000 (10 GeV/c) events have been generated for p+C
interactions,  15000 (4.5 AGeV/c), 8000 (4.5 AGeV/c) and 7230 (10 GeV/c)
events for He+Li, C and p+Ta collisions, correspondingly, by using the
enlarged UrQMD model (UrQMD+SMM). The experimental conditions and
selection criteria are applied to the generated events. For UrQMD+SMM
events the projection of the transverse momentum onto true reaction plane
was determined. The values of the flow parameter F are extracted for
protons and $\pi^{-}$ -mesons from the dependencies of $<p_{x}(y)>$ on
rapidity y for each nuclear pair (Table 1). As seen, there is a good agreement
between the experimental and theoretical values (Figs. 3, 4).

The reduced magnitude of the average pion momentum component in the reaction
plane compared to the proton one was seen before at Bevalac, GSI-SIS,
CERN-SPS and STAR \cite{R9,R23,R24}. Historically, the pattern of pion emission
relative to the reaction plane has been first studied at Bevalac by the
Streamer Chamber Group and later by the EOS collaboration \cite{R25}. In the
investigations at Bevalac and Saturne, projection of the pion transverse momentum
onto the reaction plane was examined. The direction of the pion flow opposite to
the proton flow, the so called anti-flow, was seen before in either asymmetric or symmetric systems \cite{R9,R25,R26}. However, we are unaware of observation of the pion anti-flow in a strongly asymmetric system such as our p+Ta one.

The anti-correlation of nucleons and pions was explai\-ned in \cite{R27} by the
effect of multiple $\pi$N scattering. However, in \cite{R28} it is shown that the anti-correlation is a manifestation of the nuclear shadowing of the target
and projectile spectators through both pion re-scattering and re-absorptions.
Quantitatively, the shadowing can produce the in-plane transverse momentum
components comparable to the momenta itself and, thus, much larger than
the components due to the collective motion for pions \cite{R29}. In our opinion, our
results indicate that the flow behaviour of $\pi^{-}$ mesons is a result
of the nuclear shadowing effect.

\section{Proton elliptic flow}
We have studied a proton elliptic flow in p+C (4.2 and 10
GeV/c), He+Li, C (4.5 AGeV/c) and p+Ta (10 GeV/c) collisions.
The azimuthal $\phi$ distributions of the protons were
obtained and presented in Figs 5, 6 where $\phi$ is the
angle between the transverse momentum of each particle in the
event and the reaction plane (cos$(\phi)$ = $P_{x}$
/$P^{\perp}$). The azimuthal angular distributions show maxima
at $\phi$ =90$^{\circ}$ and 270$^{\circ}$ with respect to
the event plane. The maxima are associated with a preferential
particle emission perpendicular to the reaction plane
(squeeze-out). To treat the data in a quantitative way, the
azimuthal distributions were fitted with the Fourier
cosine-expansion (given the system invariance under reflections
with respect to the reaction plane).
\begin{equation}
\frac{dN}{d\phi}=a_{0}[1+a_{1}cos(\phi)+a_{2}cos(2\phi)].
\end{equation}
The squeeze-out signature is a negative value of the coefficient $a_{2}$, which
is a measure of the strength of the anisotropic emission. The elliptic
anisotropy, quantified in terms of the $a_{2}$ coefficient ($a_{2}$=2$v_{2}$),
extracted from the azimuthal distributions of the protons with respect to the
reaction plane at mid-rapidity is given in Table 2.

The obtained experimental results have been compared with the calculations of
the UrQMD+SMM model. The experimental selection criteria were applied
to the generated events. The elliptic flow parameters with respect to the
true reaction plane were calculated (Table 2) for UrQMD+SMM events
too. A good agreement between experimental and theoretical distributions has
been obtained for the proton elliptic flow in the above mentioned collisions (Figs 5, 6).

The absolute value of the proton elliptic flow parameter
$a_{2}$ increases with the growth of momenta per nucleon in our
collisions (Table 2). According to the investigations of Au+Au
collisions at AGS \cite{R17}, the sign of the elliptic flow
changes at an apparent transition energy of E$_{tr} \sim$ 4
GeV/nucleon. All considered theoretical scenarios \cite{R34}
properly describe the change of $a_{2}$ sign at the incident
energy decrease below  $\sqrt{s_{NN}}$ = 3.5 GeV. In heavy ion
collisions the squeeze-out of particles from interaction zone
takes place due to a shadowing of particles production in the
reaction plane by the nuclear residuals. At higher energies the
shadowing decreases, and the squeeze-out flow is changed into
the elliptic flow. In the interactions studied by us (p+C and
p+Ta collisions at 4.2 and 10 GeV/c, correspondingly, there are
no projectile remnants. Thus, it is natural that the sign of
the observed elliptic flow does not change.

\section{Conclusion}
The directed transverse collective flows of protons and $\pi^{-}$ mesons and
elliptic flow of protons emitted from p+C (4.2 and 10 GeV/c), He+Li, C
(4.5 AGeV/c) and p+Ta (10 GeV/c) collisions have been studied.
In more detail:

1) The p+C system is the lightest studied one, and the p+Ta
system is an extremely asymmetrical system in which collective
flow effects (directed and elliptic) are detected for
protons and pions. As shown, the $\pi^{-}$ mesons exhibit
an opposite directed flow with that for protons in all colliding
systems. The absolute value of the directed flow parameter $F$
decreases with increase of projectile momenta in p+C collisions
for the protons from 125.2 $\pm$ 7.2 MeV/c at 4.2 GeV/c down to
85.7 $\pm$ 5.8 MeV/c at 10 GeV/c. The values for $\pi^{-}$
mesons are -21.6 $ \pm$ 11.1 MeV/c  at 4.2 GeV/c and -16.1 $
\pm$ 5.4 MeV/c at 10 GeV/c. Also, $F$ decreases with increasing
the mass numbers of the target $A_{T}$ nuclei, for protons from 85.7
$\pm$ 5.8 MeV/c (p+C, 10 GeV/c) down to 76.1 $\pm$ 5.3 MeV/c
(p+Ta, 10 GeV/c), and almost does not change for $\pi^{-}$
mesons -19.7 $ \pm$ 4.8 MeV/c (p+C, 10 GeV/c) and -18.8 $\pm$
4.8 MeV/c (p+Ta,  10 GeV/c). The results for
nucleon-nucleus collisions are opposite the results in
nucleus-nucleus interactions obtained at the same energy and on
the same experimental setup.

2) It should be mentioned that no change of the sign of the proton
elliptic flow has been observed in p+C, Ta nucleon-nucleus collisions in
the projectile momentum range of 4 $\div$ 10 GeV/c.
The absolute value of the
proton elliptic flow parameter $a_{2}$ in p+C collisions increases with projectile
momentum from -0.053 $\pm$ 0.020 at 4.2 GeV/c up to -0.071 $\pm$ 0.013 at 10 GeV/c.
Also, $a_{2}$ almost does not change with increase of the mass numbers of
the target $A_{T} $ nuclei at 10 GeV/c: -0.071 $\pm$ 0.013 (p+C) and -0.071 $\pm$ 0.016 (p+Ta).

An agreement between experimental and theoretical (UrQMD+SMM)
collective flow distributions has been obtained for particles in the
above mentioned interactions.

\section* {Acknowledgements}
\vspace{.5cm} One of us (L. Ch.) would like to thank the board of
directors of the Laboratory of Information Technologies (LIT) of
JINR for the warm hospitality.

\vspace{.5cm}
This work was partially supported by the Georgian Shota Rustaveli National
Science Foundation under Grant DI/38/6-200/13.

The authors are thankful to heterogeneous computing (HybriLIT)
team of the Laboratory of Information Technologies of JINR  for
support of our calculations.
%
%

\newpage

{\bf Figure and Table captions}

\vspace{5mm}
{\bf Fig. 1.}
Distributions of $\pi^{-}$ -mesons ($\bullet$) and $\pi^{+}$ -mesons
(after the "identification") ($\circ$), and $\pi^{+}$ -mesons (UrQMD+ SMM generated)
($\star$) in He+Li, C collisions on total momentum (P) and transverse momentum (P$_{T}$).

\vspace{5mm}
{\bf Fig. 2.}
Distributions of $\pi^{-}$ -mesons ($\bullet$), $\pi^{+}$ -mesons (before the "identification")
($\blacktriangle$),
$\pi^{+}$ -mesons (after the corrections) ($\circ$) and
$\pi^{+}$ -mesons (UrQMD+SMM generated) ($\star$) in p+Ta collisions.

\vspace{5mm}
{\bf Fig. 3.}
The dependence of $<p_{x}(y)>$ on the rapidity $y$ in p+C collisions at
the momenta of 4.2 GeV/c and of 10 GeV/c for protons and $\pi^{-}$
mesons in experimental ($\bullet$, $\blacktriangle$) and UrQMD+SMM generated
($\circ$, $\triangle$) data, correspondingly.
Straight solid lines stretches represent the slope of data at
mid-rapidity,
obtained by fitting the data with a 1-st order polynomial within the
intervals of the rapidity. The curved lines guide the eye over data.
Arrows indicate average y$_{c}$ over the interactions.

\vspace{5mm}
{\bf Fig. 4.}
The dependence of $<p_{x}(y)>$ on the rapidity $y$ in He+Li,C
collisions at 4.5 AGeV/c and p+Ta interactions at 10 GeV/c for
protons and $\pi^{-}$ -mesons in experimental
($\bullet$, $\blacktriangle$) and
UrQMD+SMM generated ($\circ$, $\triangle$) data,
correspondingly.
Arrows indicate average y$_{c}$ over the interactions.

\vspace{5mm}
{\bf Fig. 5.}
The azimuthal distributions with respect to the reaction plane in p+C
collisions at the momenta of 4.2 GeV/c and of 10 GeV/c for protons.
The curves are the result of the approximation by
$dN/d\phi=a_{0}[1+a_{1}cos(\phi)+a_{2}cos(2\phi)]$.

\vspace{5mm}
{\bf Fig. 6.}
The azimuthal distributions with respect to the reaction plane in He+Li, C
collisions at 4.5 AGeV/c and p+Ta interactions at 10 GeV/c for protons
in experimental ($\bullet$) and UrQMD+SMM generated ($\circ$) data,
correspondingly. The curves are the result of the approximation by
$dN/d\phi=a_{0}[1+a_{1}cos(\phi)+a_{2}cos(2\phi)]$.

\vspace{5mm}
{\bf Table 1.}
 The numbers of experimental and UrQMD+\\ SMM events prior and after
applying the cuts (see text) and the values of flow parameters for
protons and $\pi^{-}$ -mesons.

\vspace{5mm}
{\bf Table 2.}
 Characteristics of proton elliptic flow in the experimental
and UrQMD+SMM collisions including event number ($N_{exp.}$, $N_{UrQMD}$)
prior and after applying the cuts.

\newpage

\onecolumn

\begin{figure}
\begin{center}
\includegraphics[width=70mm,height=70mm,clip]{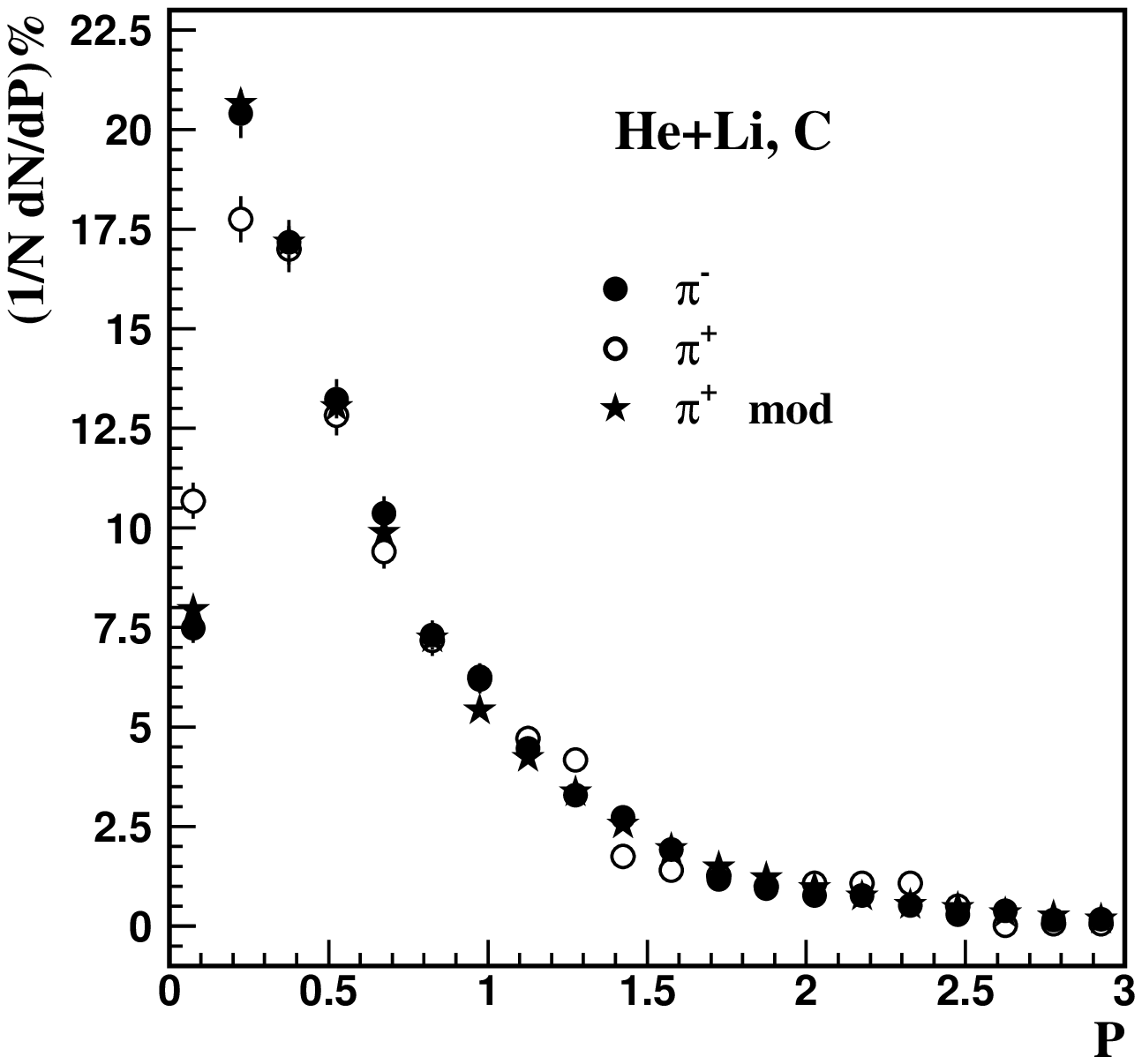}
\hspace{1cm}
\includegraphics[width=70mm,height=70mm,clip]{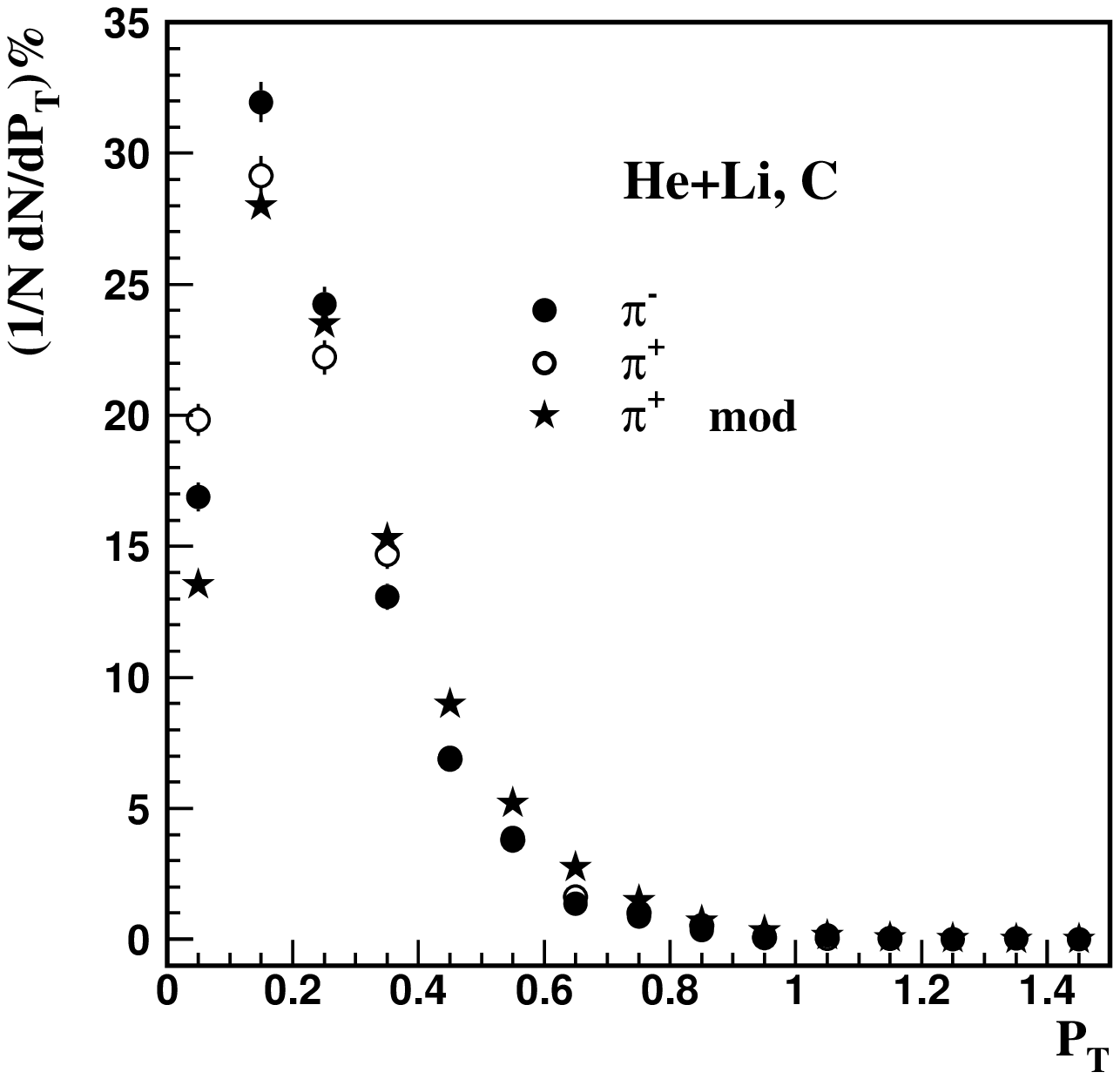}
\caption{}
\end{center}
\end{figure}

\begin{figure}[cbth]
\vspace{3cm}
\begin{center}
\includegraphics[width=70mm,height=70mm,clip]{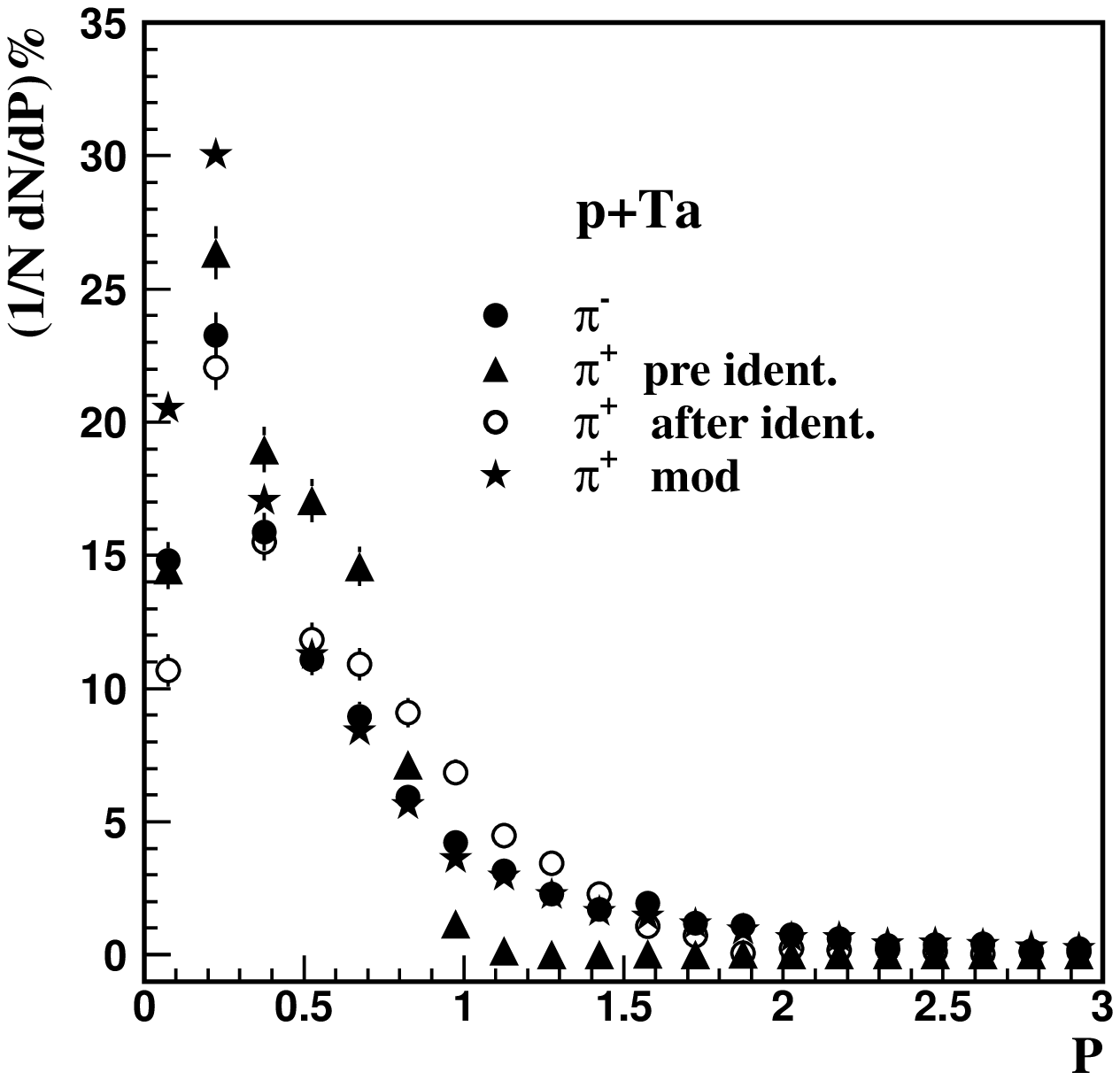}
\hspace{1cm}
\includegraphics[width=70mm,height=70mm,clip]{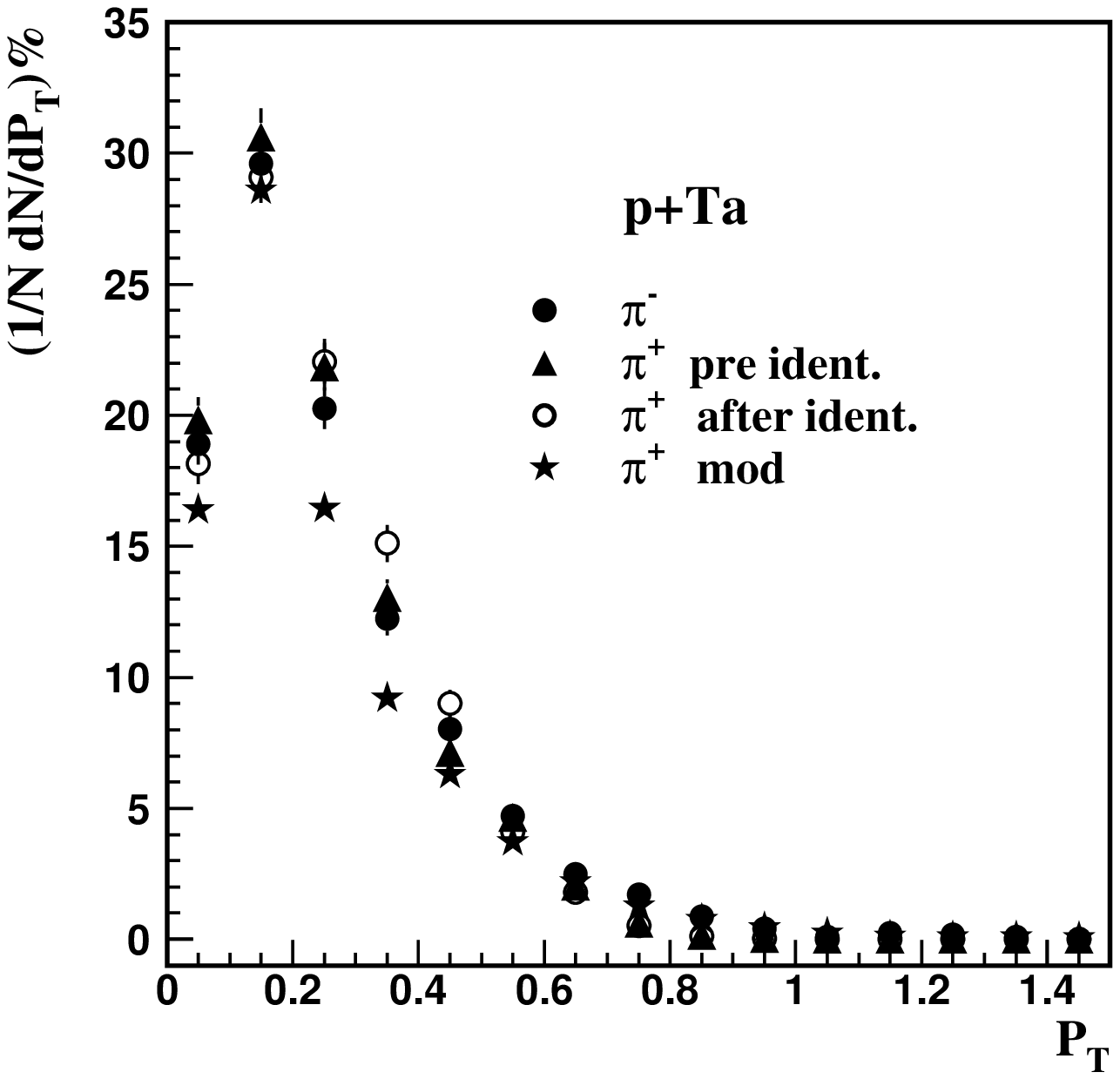}
\caption{}
\end{center}
\end{figure}

\newpage

\begin{figure}[cbth]
\begin{center}
\includegraphics[width=80mm,height=80mm,clip]{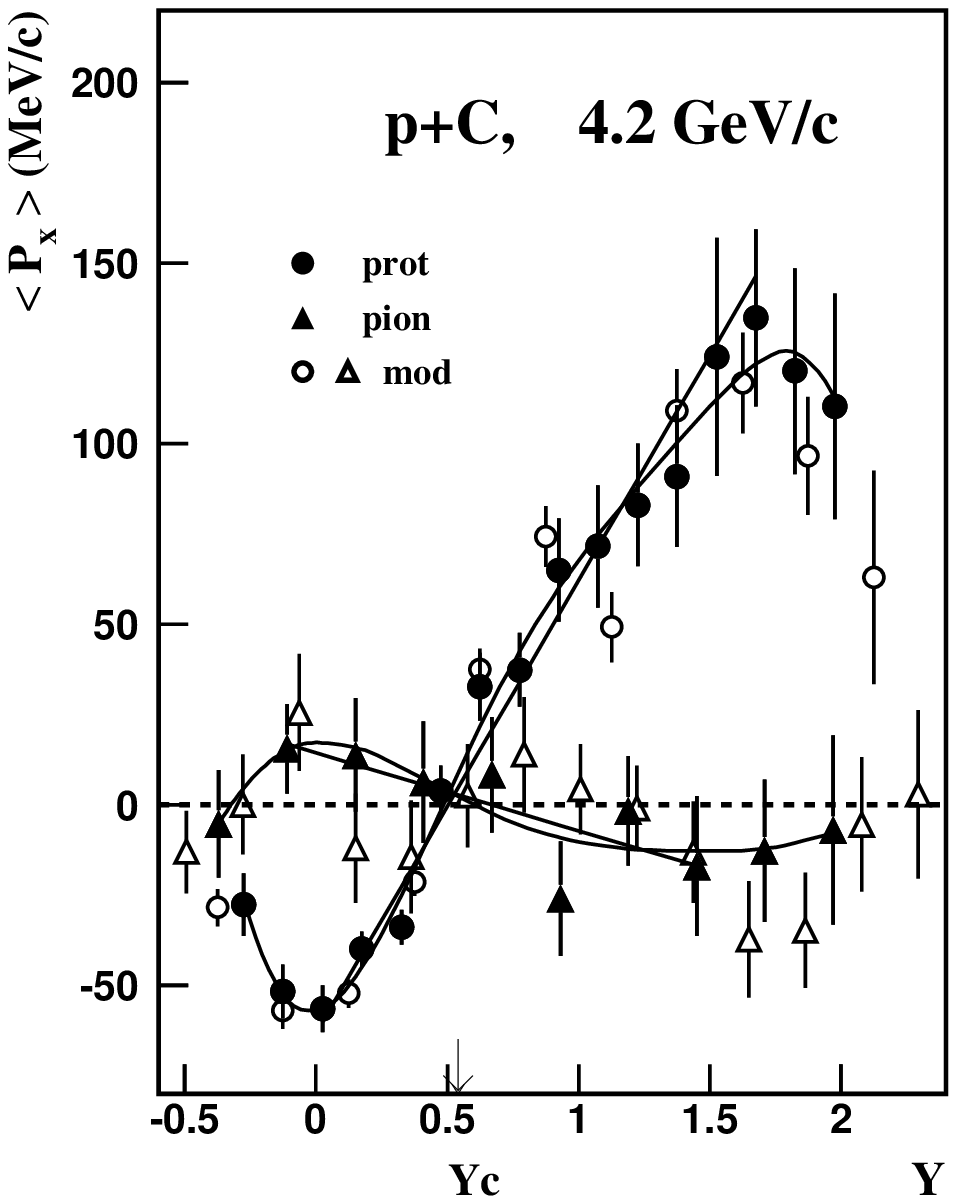}
\hspace{1cm}
\includegraphics[width=80mm,height=80mm,clip]{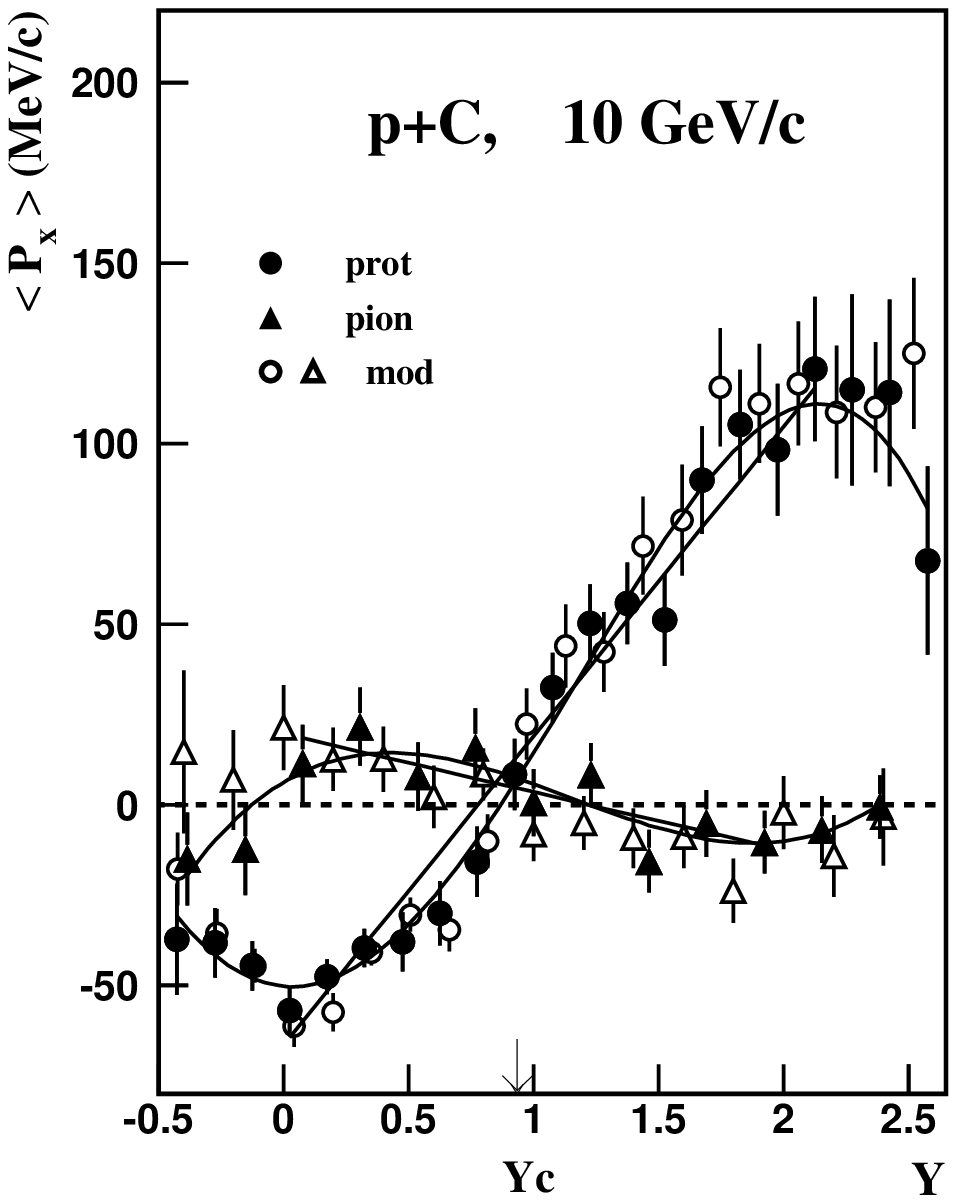}
\caption{}
\end{center}
\end{figure}

\vspace{3cm}

\begin{figure}[cbth]
\begin{center}
\includegraphics[width=70mm,height=70mm,clip]{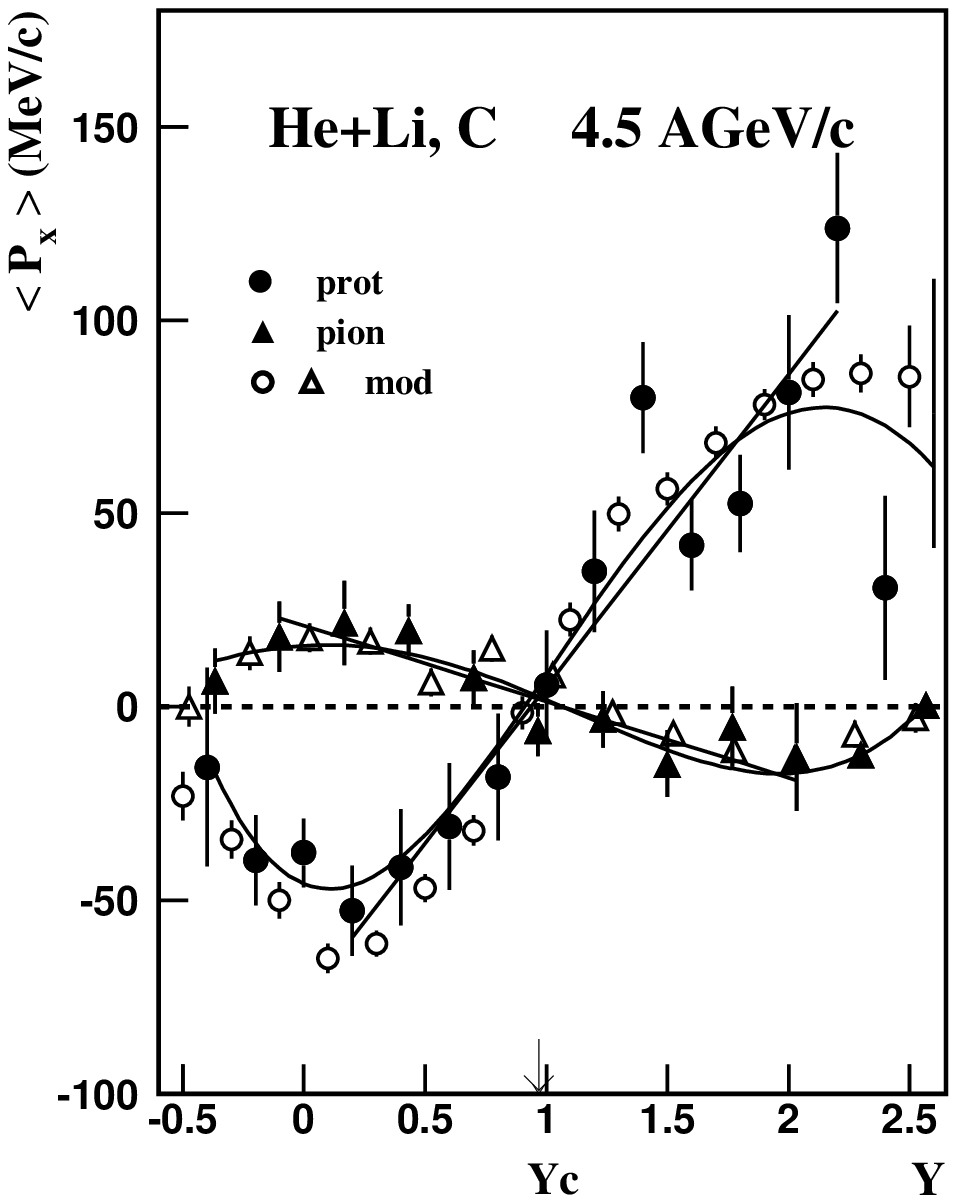}
\hspace{1cm}
\includegraphics[width=70mm,height=70mm,clip]{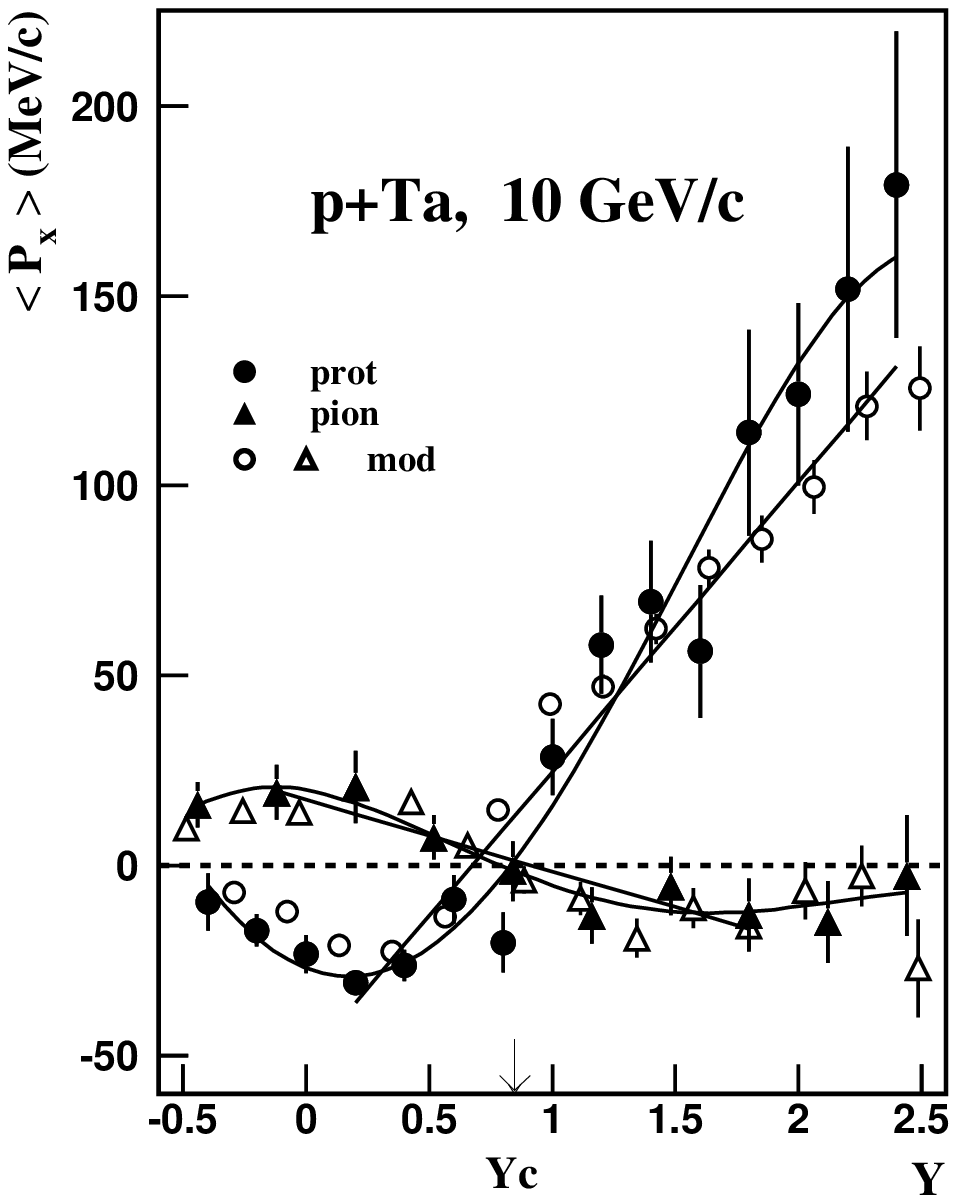}
\caption{}
\end{center}
\end{figure}

\newpage

\begin{figure}[cbth]
\begin{center}
\includegraphics[width=70mm,height=70mm,clip]{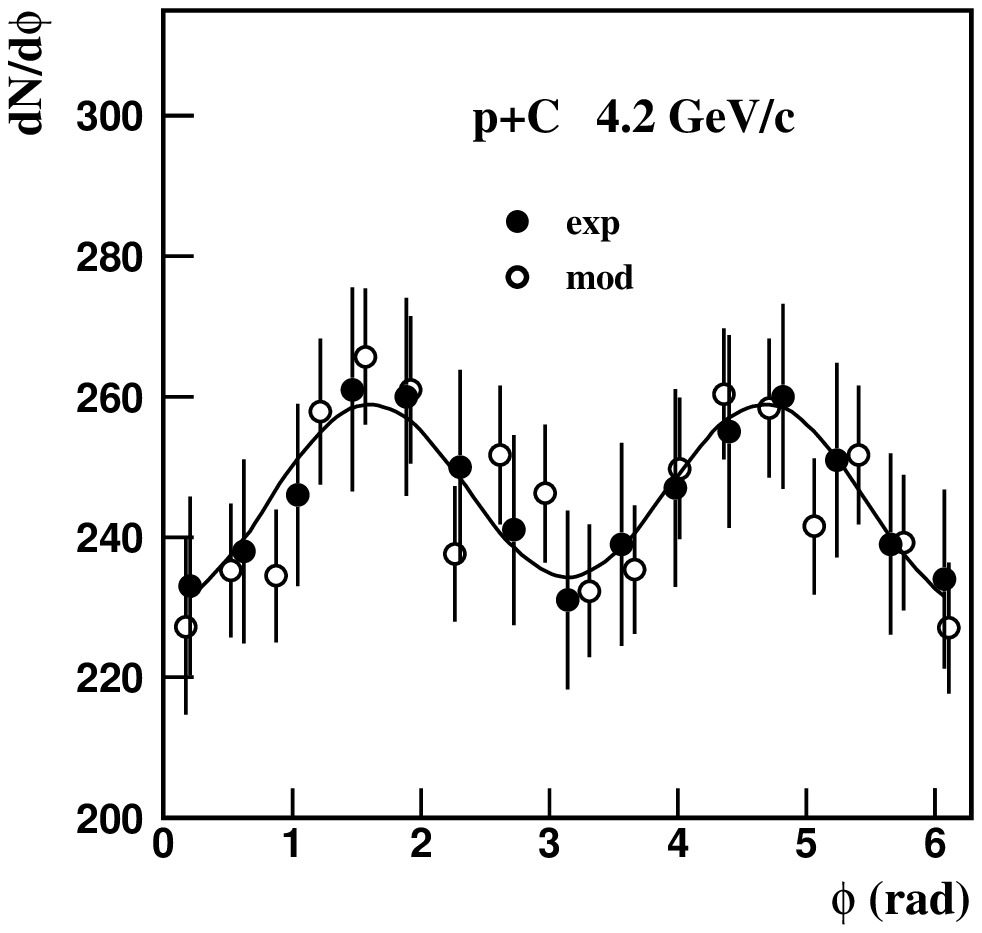}
\hspace{1cm}
\includegraphics[width=70mm,height=70mm,clip]{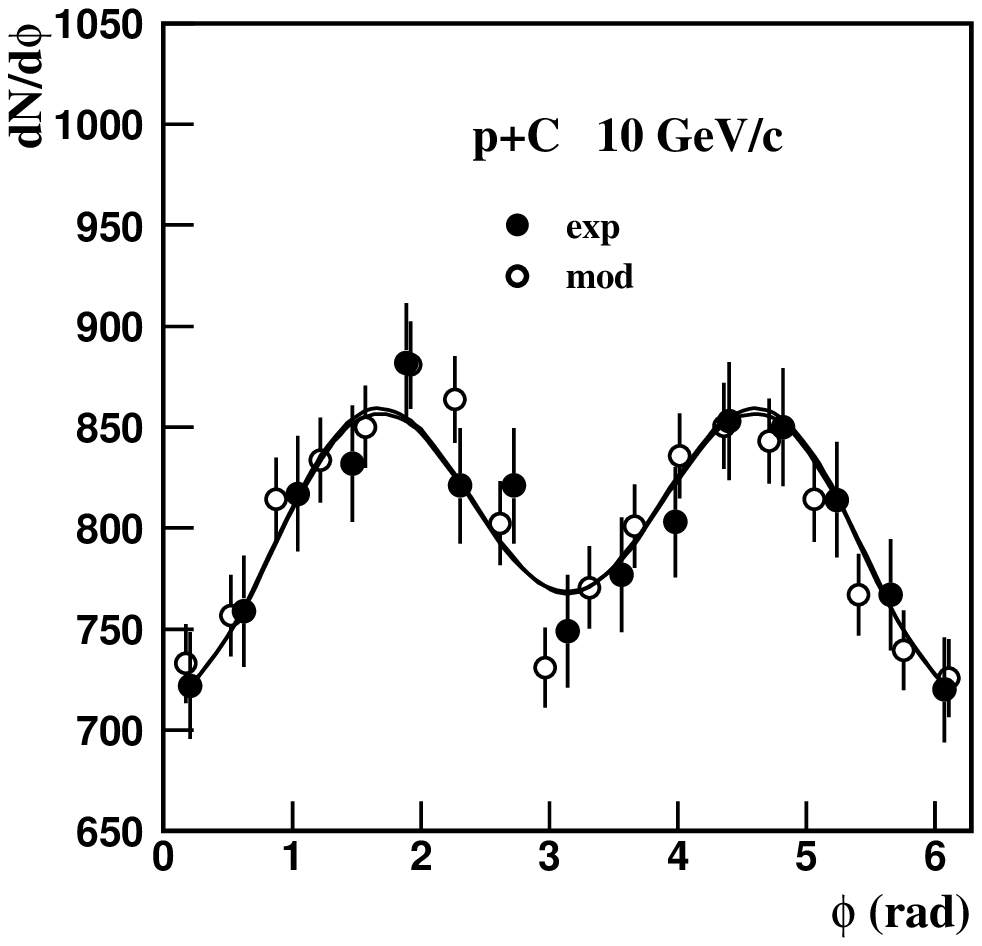}
\caption{}
\end{center}
\end{figure}

\vspace{3cm}

\begin{figure}[cbth]
\begin{center}
\includegraphics[width=80mm,height=80mm,clip]{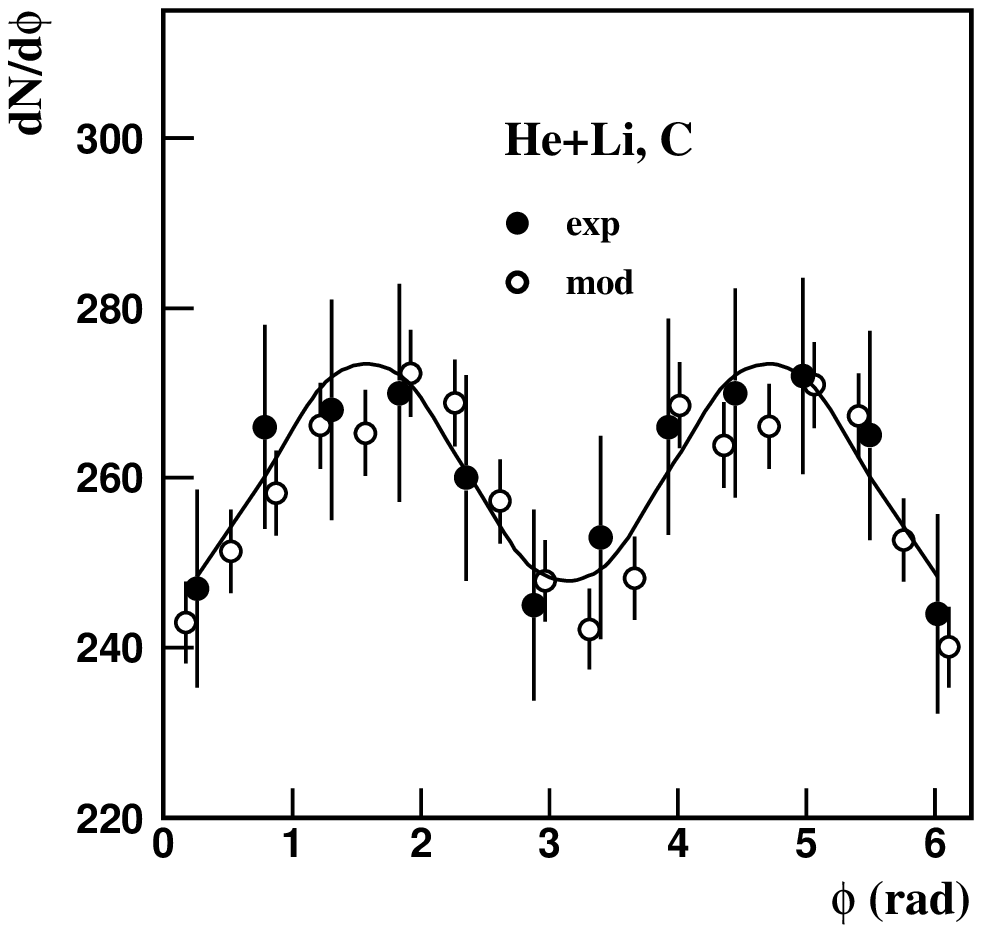}
\hspace{1cm}
\includegraphics[width=80mm,height=80mm,clip]{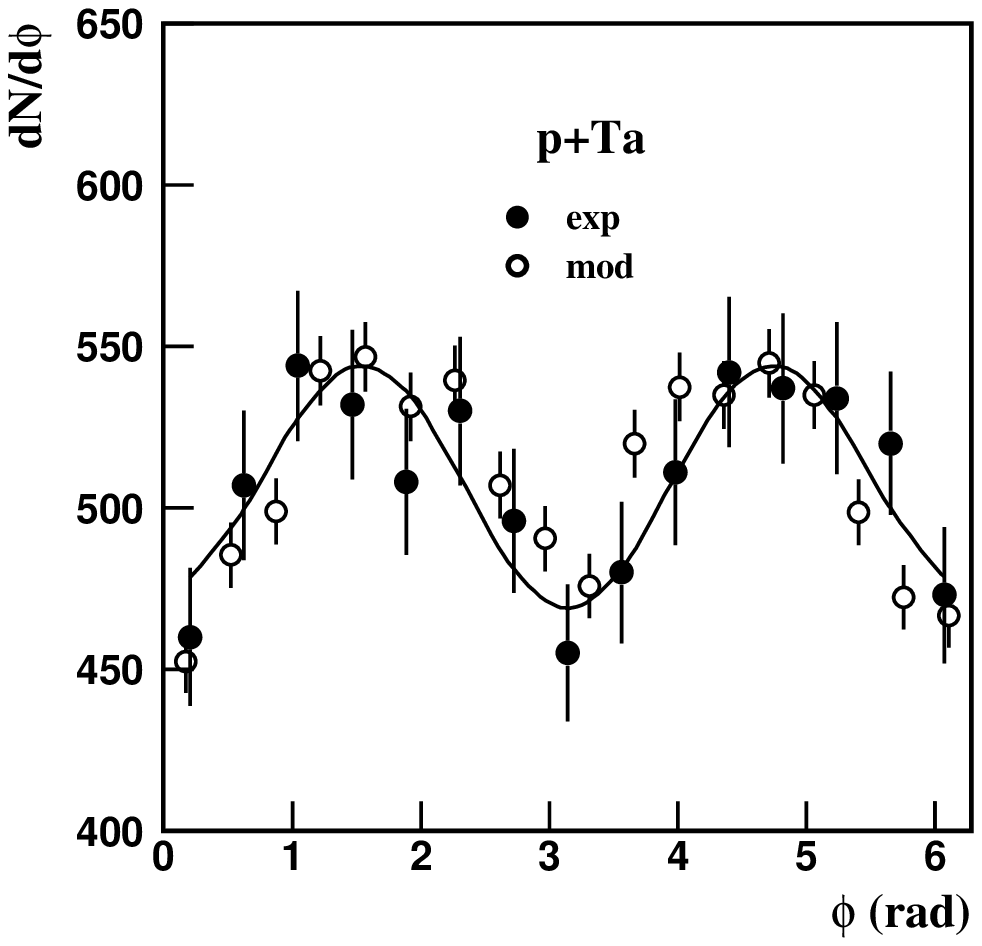}
\caption{}
\end{center}
\end{figure}

\begin{table}
\caption{}
\begin{center}
\begin{tabular}{|c|c|c|c|c|}
\hline
&    &     &    &       \\
$A_{P} + A_{T}$ & p+C & p+C & He+Li, C &   p+Ta    \\
& 4.2 GeV/c  & 10 GeV/c    &      &         \\
&    &     &    &       \\
\hline
&    &     &    &       \\
$N_{\rm{prior\hspace{0.1cm} exp.}}$ & 5882 & 16509 & 6147& 2342\\
 $N_{\rm{after\hspace{0.1cm} cut }}$ & 891 & 2890 & 786 &  1141\\
&    &     &    &       \\
\hline
&    &     &   &         \\
$N_{\rm{prior\hspace{0.1cm} mod.}}$ &20000 & 60000 & 23000 & 7230 \\
 $N_{\rm{after \hspace{0.1cm} cut }}$ & 1428  & 4858 & 10435 & 6333 \\
&    &     &   &         \\
\hline
&    &     &   &         \\
$<cos\Phi>$  exp. & 0.633 & 0.638 & 0.662 & 0.642 \\
&    &     &   &  \\
\hline
&    &     &   &         \\
$<cos\Phi>$  mod. & 0.657 & 0.658 & 0.670 &  0.702 \\
&    &     &   &  \\
\hline
&    &     &   &    \\
$F^{\rm{p}}_{\hspace{0.1cm} exp.}$(MeV/c)& 125.2 $\pm$ 7.2& 85.7 $\pm$ 5.8
& 80.9 $\pm$ 16.2 & 76.1 $\pm$ 5.3 \\
&    &     &   &   \\
\hline
&    &     &   &         \\
$F^{\rm{p}}_{\hspace{0.1cm} mod.}$ (MeV/c) & 116.2 $\pm$ 4.9 & 94.3 $\pm$ 3.7
& 86.2 $\pm$ 1.9 & 79.8 $\pm$ 1.9 \\
&    &     &   &  \\
\hline
&    &     &   &         \\
$F^{\rm{\pi^{-}}}_{\hspace{0.1cm} exp.}$(MeV/c) & -21.6 $\pm$ 11.1 & -16.1 $\pm$ 5.4
& -19.7 $\pm$ 4.8 & -18.8 $\pm$ 4.8 \\
&    &     &   &  \\
\hline
&    &     &   &    \\
$F^{\rm{\pi^{-}}}_{\hspace{0.1cm} mod.}$ (MeV/c)  & -19.0 $\pm$ 7.9 & -15.9 $\pm$ 3.9
& -17.2 $\pm$ 1.8 & -18.1 $\pm$ 1.9 \\
&    &     &   &  \\
\hline
\end{tabular}
\end{center}
\end{table}

\begin{table}
\vspace{3cm}
\caption{}
\begin{center}
\begin{tabular}{|c|c|c|c|c|}
\hline
&    &     &    &        \\
  & $N_{\rm{prior\hspace{0.1cm} exp.}}$    & $N_{\rm{prior\hspace{0.1cm}
mod.}}$  & &  \\
$A_{P} + A_{T}$ &      &      & $a_{\rm{2 \hspace{0.1cm} exp.}}$ &
$ a_{\rm{2 \hspace{0.1cm} mod.}}$  \\
& $N_{\rm{after \hspace{0.1cm}cut }}$ & $N_{\rm{after
\hspace{0.1cm} cut }}$& & \\
&    &     &   &         \\
\hline
4.2 GeV/c & 5882 & 20000 &           &
\\
          & 891 &  1428 & -0.053 $\pm$ 0.020    & -0.052 $\pm$ 0.013
\\
p+C       &    &     &   &         \\
10 GeV/c  & 16509 & 60000 &                       &
\\
          & 2890  & 4858 &  -0.071 $\pm$ 0.013  & -0.072 $\pm$ 0.008
\\
\hline
          & 6147  & 23000 &    &             \\
 He+Li, C     &       &       & -0.051 $\pm$ 0.019    & -0.052 $\pm$ 0.007
\\
          & 786  &  10435 &    &            \\
\hline
          & 2342  & 7230 &    &            \\
 p+Ta      &       &       & -0.071 $\pm$ 0.016 & -0.072 $\pm$ 0.007     \\
          & 1141  & 6333 &    &            \\
\hline
\end{tabular}
\end{center}
\end{table}

\end{document}